\documentclass[12pt]{iopart}
\usepackage{iopams}
\usepackage{graphicx}
\usepackage{color}
\usepackage{amsfonts}
\input amssym.def 
\input amssym

\begin{document}

\title[]{Geometry of contextuality from Grothendieck's coset space}

\author{ Michel Planat
}

\vspace*{.1cm}
\address
{
 Institut FEMTO-ST, CNRS, 15 B Avenue des Montboucons, F-25044 Besan\c con, France. ({\tt michel.planat@femto-st.fr})}

\vspace*{.2cm}






\begin{abstract}\\
The geometry of cosets in the subgroups $H$ of the two-generator free group $G=<a,b>$ nicely fits, via Grothendieck's 
dessins d'enfants, the geometry of commutation for quantum observables. Dessins stabilize point-line incidence geometries that reflect the commutation of (generalized) Pauli operators [Information 5, 209 (2014); 1310.4267 and 1404.6986 (quant-ph)]. Now we find that the non-existence
 of a dessin for which the commutator $(a,b)=a^{-1}b^{-1}ab$ precisely corresponds to the commutator of quantum observables $[\mathcal{A},\mathcal{B}] = \mathcal{A}\mathcal{B}-\mathcal{B}\mathcal{A}$ on all lines of the geometry is a signature of quantum contextuality. 
This occurs first at index $|G:H|=9$ in Mermin's square and at index $10$ in Mermin's pentagram, as expected. Commuting sets of $n$-qubit observables with $n>3$ are found to be contextual as well as most generalized polygons. A geometrical contextuality measure is introduced. 

\end{abstract}

\vspace*{-.5cm}
\normalsize

\section{Introduction}
\noindent

{\it Never ask for the meaning of a word in isolation, but only in the context of a sentence} (Gottlob Frege, Grundlagen der Arithmetik, 1884).

 {\it There is no quantum world. There is only an abstract quantum physical description. It is wrong to think that the task of physics is to find out how Nature is. Physics concerns what we say about Nature} [Niels Bohr, Spoken at the Como conference, 1927].

The lack of commutativity of quantum observations gives rise to the concept of contextuality, a kind of impossibility to recover the quanta of reality irrespectively of our words for describing it. In a nutshell, Kochen-Specker theorem states that contextuality is needed to reproduce all quantum mechanical predictions on a $d$-dimensional ($d>3$) system \cite{Kochen1967}. Since this foundational no-go theorem was discovered, many quantum systems carrying quantum contextuality have been displayed, see \cite{Waegell2014,Howard2013,Spekkens2014} for a recent hint. One of the most transparent contextuality proofs consist of particular sets of observables in a four-dimensional (two-qubit) or in a eight-dimensional (three-qubit) system, through the geometries of Mermin's square and pentagram, respectively \cite{Mermin1993,Planat2013bis}. Contextuality in such systems was shown to be experimentally testable \cite{Cabello2008}.

We recently found a mathematical scheme giving rise to the aforementioned geometries as well as many related ones \cite{PlanatFQXi}-\cite{Dessins2014}. Our work is based on Grothendieck's great insight about the relationship between algebra, geometry and topology, called by him {\it dessins d'enfants} \cite{Groth84} \footnote{For another mathematical approach of quantum contextuality based on sheaf theory, the reader should look at \cite{Abramsky2011} and the references therein.}. In the present note, the elusive \lq contextual' geometries are given a precise definition. We compare the (ring) commutativity of observables and the (group theoretical) commutativity of cosets that both coordinatize the vertices of the relevant geometry. We find that the non-existence of a dessin for which the coset commutator $(a, b) = a^{-1}b^{-1}ab$ exactly corresponds to the commutator of observables
$[\mathcal{A},\mathcal{B}] = \mathcal{A}\mathcal{B}-\mathcal{B}\mathcal{A}$ on all lines of the geometry is a convincing signature of quantum contextuality. In this definition, contextuality arises first for $9$ and $10$ vertices like in Mermin's point-line configurations.

In Sec. 2, we shortly explain how the two-generator free group and its subgroups are given the coset structure of a Grothendieck's dessin d'enfant $\mathcal{D}$, how a $\mathcal{D}$ may stabilize a point/line geometry  $\mathcal{G}$, then we introduce our criterion for geometrical contextuality. In Sec. 3, we fully explicit the algebraic/topological/geometrical  meaning of small non-trivial dessins in relation to their (non-)contextuality, including the case of Mermin's structures. In Sec. 4, it is shown how contextuality arises in maximum sets of commuting operators starting with the $4$-qubit case.  
Finally, in Sec. 5, a geometrical contextuality measure is introduced and applied to generalized polygons.

\section{Coset coordinates, dessins d'enfants and finite geometries}
\noindent

Let $F=\left\langle a,b \right\rangle$ be the free group on two generators. Elements in the group are words $u$, that are products of elements of $F$ and their inverses  modulo the only defining relation $uu^{-1}=e$, with $e$ the identity element. In the following, we restrict to the free group $G=\left\langle a,b~|~ b^2=e \right\rangle$, which accounts for an extra involution $b$. The index $n:=|G:H|$ of a subgroup $H$ in $G$ counts the number of cosets/copies of $H$ that fill up $G$. A right coset with respect to an element $g \in G$ is defined as $Hg=\left\{hg:h \in H\right\}$. The set of right cosets partitions $G$. In other words, every $g \in G$ belongs to just one right coset. Similar statements holds for left cosets.

A transversal is an indexed set of (right) coset representatives for $H$ in $G$, and the coset table is a way to express the action of generators $a$, $b$ and of the non-trivial inverse $a^{-1}$ on them. The algorithm performing this task is the Coxeter-Todd algorithm \cite{Magma}. Under the action of $a$ and $b$, the indexed coset representatives are represented by a two-generator permutation group $P=\left\langle g_0,g_1 \right\rangle$. The latter corresponds to a map on a compact orientable surface that is, a triple $(g_0,g_1,g_{\infty})$ with $g_0 g_1g_{\infty}=1$, from which the $V$ vertices, $E$ edges and $F$ faces of the map are defined by the cycles of $g_0$, $g_1$ and $g_{\infty}$ \cite{Jones78,Lando2004}. Grothendieck was enthusiastic in seing such a map as a bicolored map $\mathcal{D}$, also called an hypermap \cite{Walsh}, with $B$ black vertices and $W$ white vertices, in such a way that the adjacent vertices have always opposite color and the corresponding segments are the $n$ edges \cite{Groth84}. For bicolored maps derived from $G$, the valency of white vertices  is $ \le 2$. The resulting {\it dessin d'enfant} is endowed a natural topological structure with Euler characteristic $2-2g=B+W+F-n$, where $g$ stands for the topological genus.

Grothendieck also recognized a dessin as an object defined over the field $\bar{\mathbb{Q}}$ of algebraic numbers as a complex algebraic curve. Technically, given  $f(x)$, a rational function of the complex variable $x$, a {\it critical point} of $f$ is a root of its derivative and a {\it critical value} of $f$ is the value of $f$ at the critical point. 
Let us define a so-called {\it Belyi function} corresponding to a
dessin $\mathcal{D}$ as a rational function $f(x)$ of degree $n$ embedded into
the Riemann sphere $\hat{\mathbb{C}}$ in such a way that (i) the black vertices are the roots of the equation $f(x)=0$ with the multiplicity of each root being equal to the degree of the corresponding (black) vertex, (ii) the white vertices are the roots of the equation $f(x)=1$ with the multiplicity of each root being equal to the degree of the corresponding (white) vertex,
(iii) the bicolored map is the preimage of the segment $[0,1]$, that is $\mathcal{D}=f^{-1}([0,1])$,
(iv) there exists a single pole of $f(x)$, i.\,e. a root of the equation $f(x)=\infty$, at each face, the multiplicity of the pole being equal to the degree of the face, and
(v) besides $0$, $1$ and $\infty$, there are no other critical values of $f$. In addition, the coefficients of Belyi functions are algebraic numbers \cite{Lando2004}. 

Finally, the coset structure and its permutation representation (by a dessin d'enfant) provide a coordinatization to many point-line geometries occuring in the investigation of commutation of quantum observables \cite{PlanatFQXi}-\cite{Dessins2014}. Taking the permutation group $P$ (it identifies a dessin $\mathcal{D}$) corresponding to a subgroup $H$ of $G$, one proceeds by first listing the $m$ non-isomorphic subgroups $S_m$ stabilizing a pair of elements/cosets. Given a $S_m$, all points on a line of the putative geometry $\mathcal{G}_m$ should share the same two-point stabilizer subgroup of $P$. The lines of a $\mathcal{G}_m$ are thus distinguished by their (isomorphic) stabilizers acting on different $G$-sets. Doing this, the cosets happen to coordinatize the edges of the $\mathcal{D}$ and, at the same time, the vertices  of the  resulting geometries $\mathcal{G}_m$.

\subsection*{Identifying commutation for cosets and for observables: contextuality}
\noindent

The key point, not recognized by us before, is that, not only there should exist a bijection between a point-line geometry $\mathcal{G}_m$ stabilized by a dessin $\mathcal{D}$ and the point-line geometry occuring in a set of quantum observables (quantum observables as cosets), but the commutation structure in both cases should also correspond (commuting operators on a line as commuting cosets). While the commutator $[\mathcal{A},\mathcal{B}] = \mathcal{A}\mathcal{B}-\mathcal{B}\mathcal{A}$ for observables $\mathcal{A}$ and $\mathcal{B}$ is that for a ring, the commutator for the representative of two cosets $a$ and $b$  is the group theoretical one $(a,b)=a^{-1}b^{-1}ab$. We identify a contextual geometry as one where at least one line of $p$ points/cosets fails to satisfy the \lq commutation law' $(a_1,a_2,\ldots,a_p)=e$ whatever the ordering of cosets. Observe that, in our definition of non-contextuality, we do not ask the commutation of all pairs of cosets but that of their product.

A few non-contextual and contextual geometries are described at the next section.  The smallest index contextual geometry is recognized to be a $3 \times 3$-grid (a Mermin square of observables) as it could have been expected.

\section{From non-contextual to contextual point-line geometries}
\noindent

\subsection*{Recovering the octahedron}
\noindent

Let us apply our approach to a very simple geometry, that of the octahedron $\mathcal{O}$ whose \lq lines' are the triangles. There exist $56$ subgroups of index $6$ in $G$ and, while many of the corresponding dessins may be used to recover $\mathcal{O}$, most of them are of the contextual type.

\begin{figure}[ht]
\centering 
\includegraphics[width=7cm]{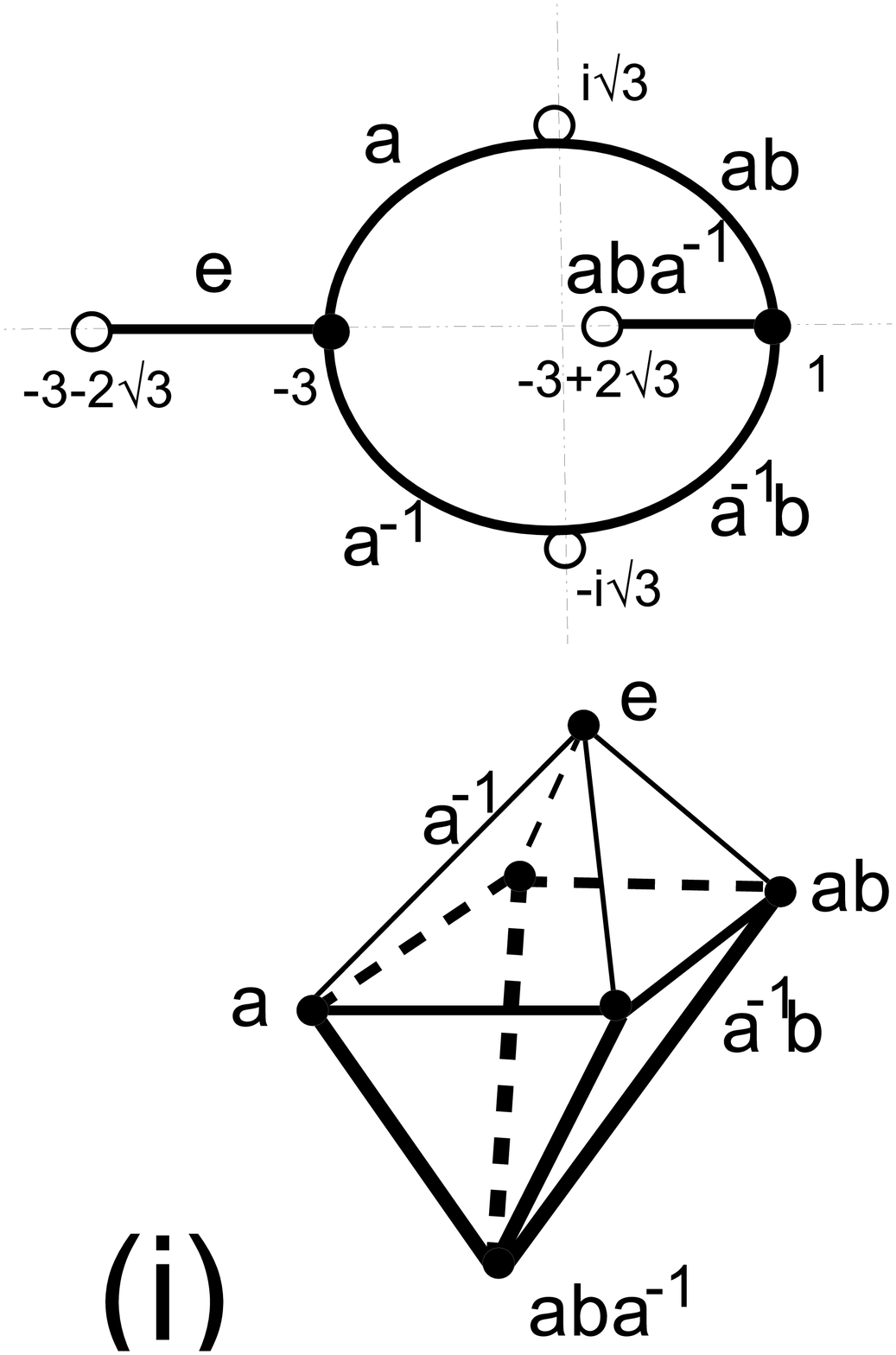}
\includegraphics[width=7cm]{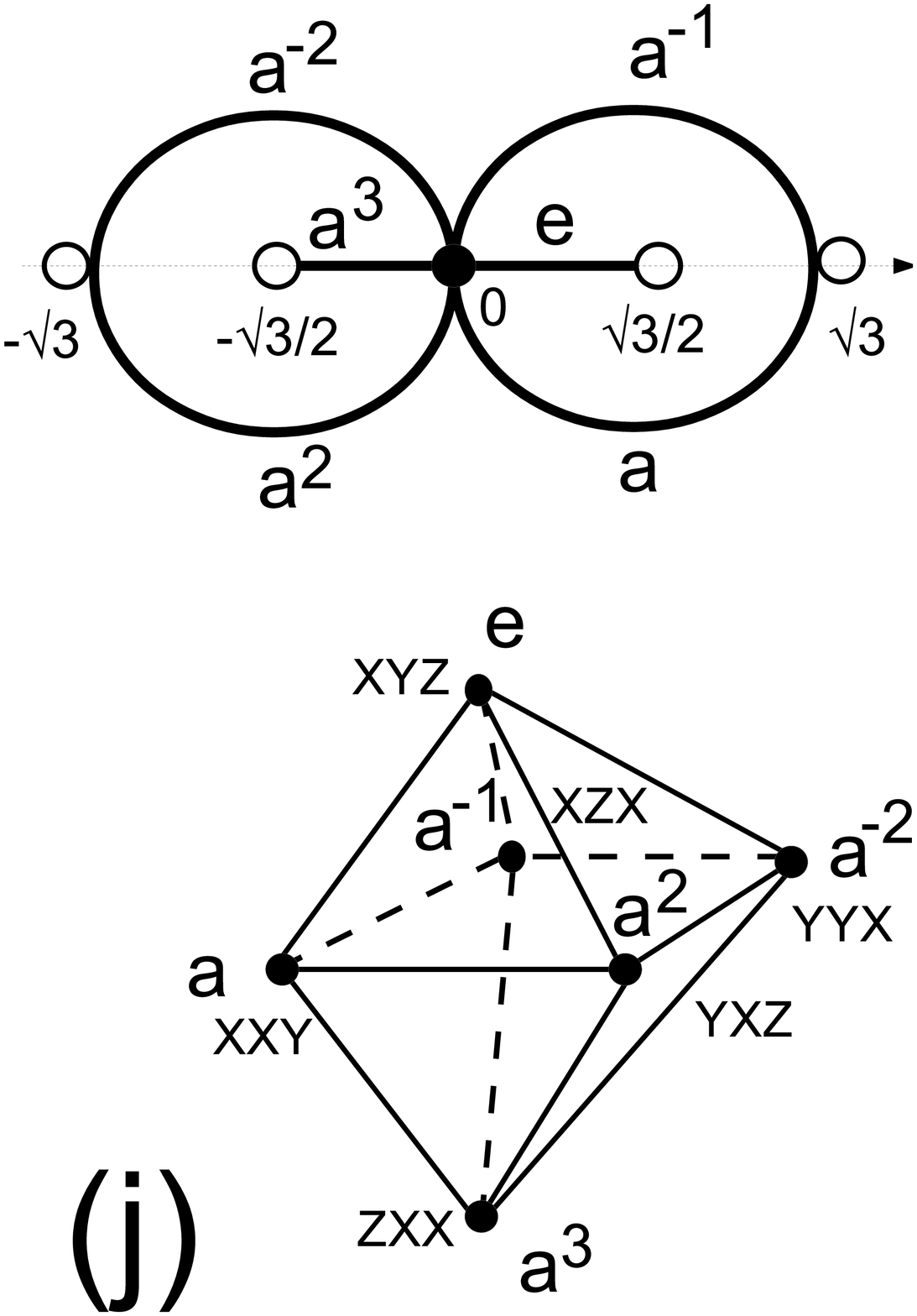}
\caption{(i) A contextual hypermap (top) stabilizing the octahedron (bottom) with the corresponding coset labelling.  The vertices of the dessin are in the extension field $\mathbb{Q}(\sqrt{3})$ as shown. The triangles with thick lines do not have all their edges indexed with commuting cosets. (j) A non-contextual map (top) stabilizing the octahedron (bottom). The octahedron is also given a set $3$-qubit coordinates that are mutually commuting at the vertices of a triangle.}
\label{fig1}  
\end{figure}

For instance, the permutation group $P_1=\left\langle (1,2,3)(4,5,6),(2,4)(3,5)\right\rangle$ [where the $G$-set $\{1,2,3,4,5,6\}$ is an ordered set of indices for the transversal $\{e,a,a^{-1},ab,a^{-1}b,aba^{-1}\}$] can be used to recover $\mathcal{O}$ as shown in Fig. 1 (i). The cosets that serve as coordinates of the edges of the dessin and as coordinates of the vertices of $\mathcal{O}$ are shown. In this setting, only the triangles above the square in $\mathcal{O}$ have their coordinates satisfying the commutation law so that the dessin is of the contextual type.

Since the dessin for $P_1$ is of a small size, it is an easy task to derive its corresponding Belyi curve as
$$f_1(x)=-\frac{1}{64x^3}(x-1)^3(x+3)^2.$$
To see this, take the derivative $f_1'(x)=3(x-1)^2(x+3)^2(x^2+3)/(64x^3)$. The critical points are the black points at $x=1$ and $x=-3$ ( where the valency is $3$) and the white points at $x= \pm i \sqrt{3}$ (where the valency is $2$). Then, solving the equation $f_1(x)=0$, one gets of course the two solutions $x=1$ and $-3$ corresponding to the black points while solving for $f_1(x)=1$ one gets the critical (white) points at $x= \pm i \sqrt{3}$ and the real (white) points at $x=-3\pm 2\sqrt{3}$. Hence the coordinates of vertices of the dessin are those shown in Fig. 1 (i). The two solutions of the equation $f_1(x)=\infty$ are $x=0$ and $x=\infty$, they correspond to the center of the faces.

\begin{figure}[ht]
\centering 
\includegraphics[width=6cm]{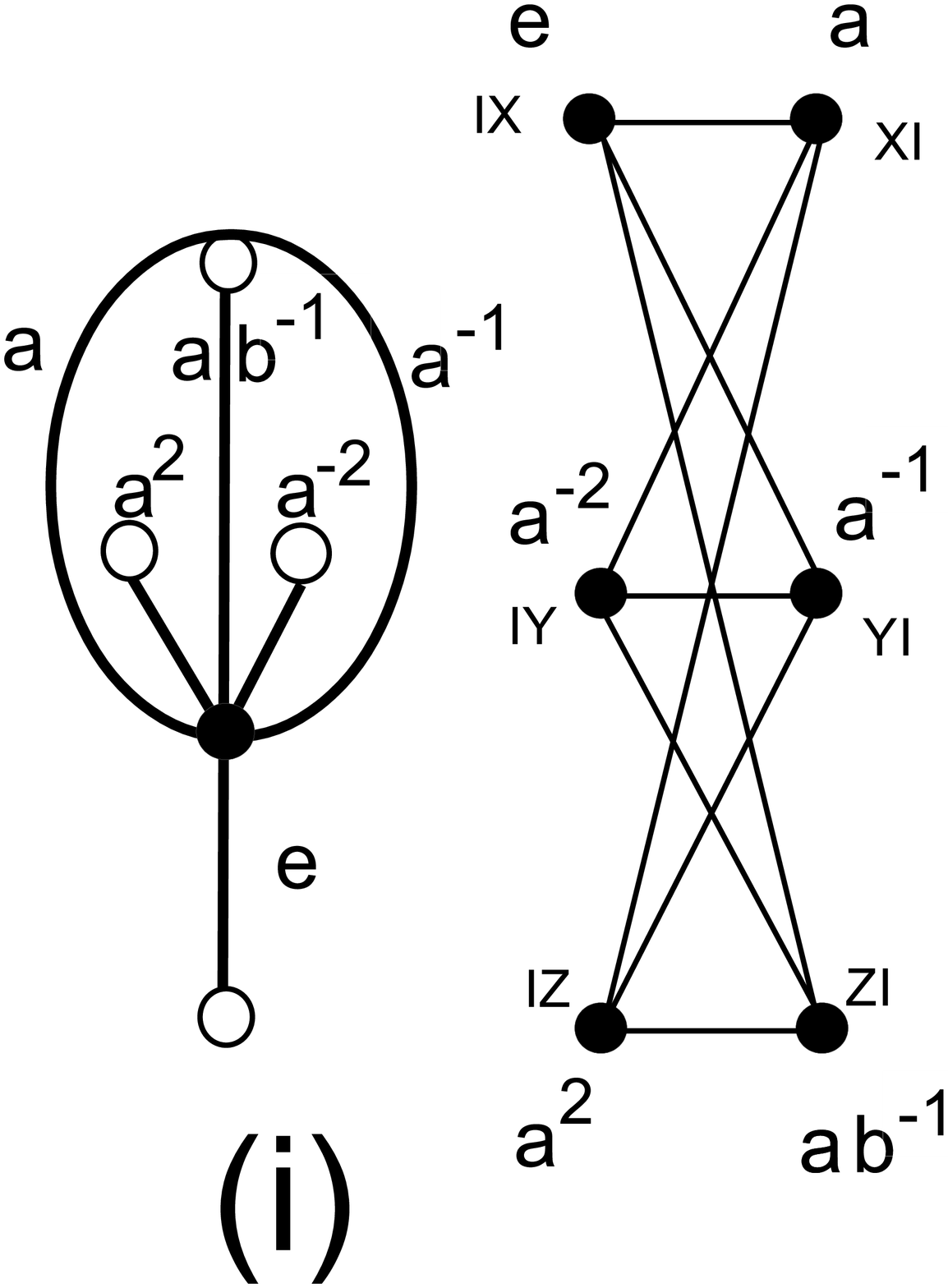}
\includegraphics[width=6cm]{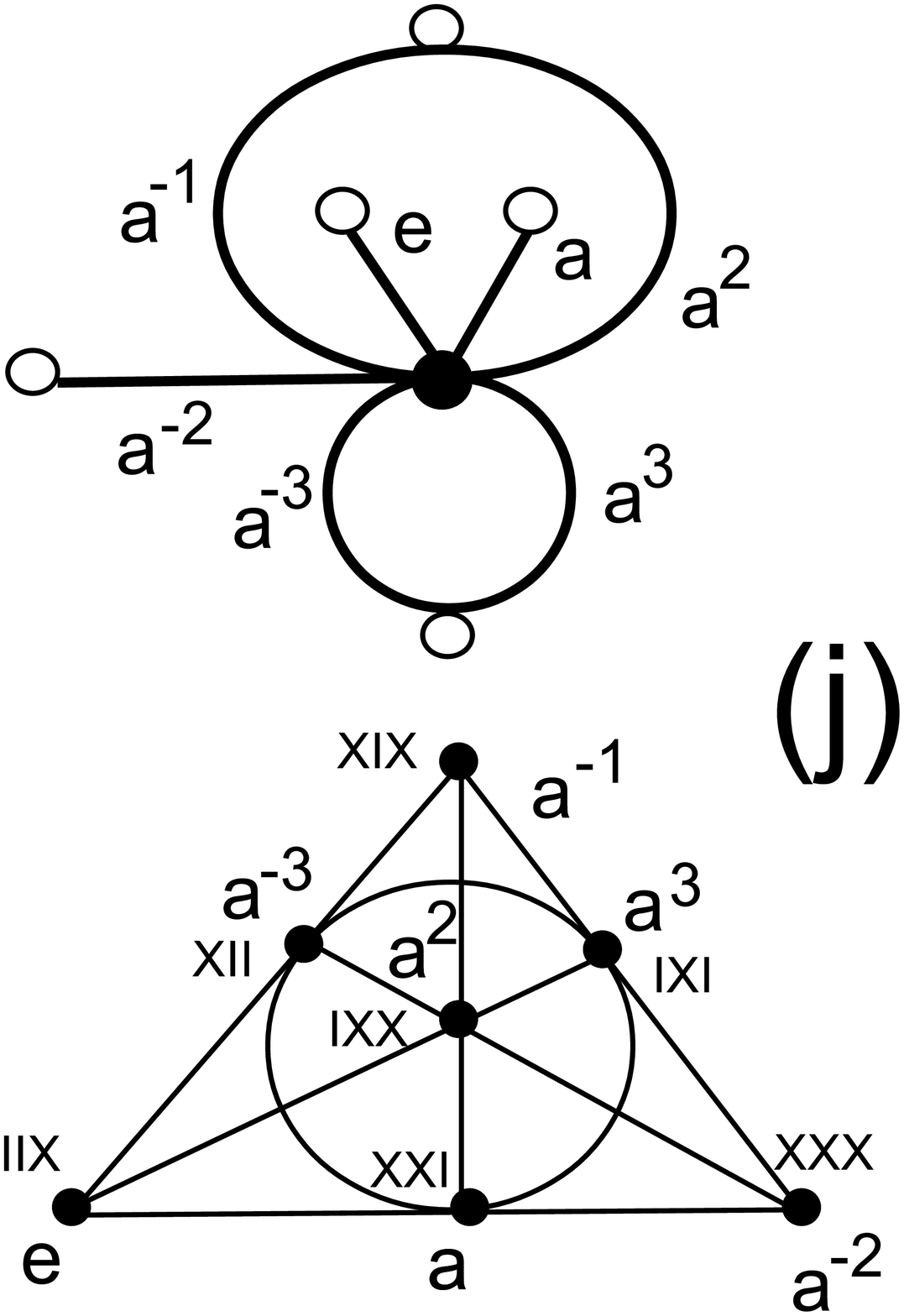}
\includegraphics[width=6cm]{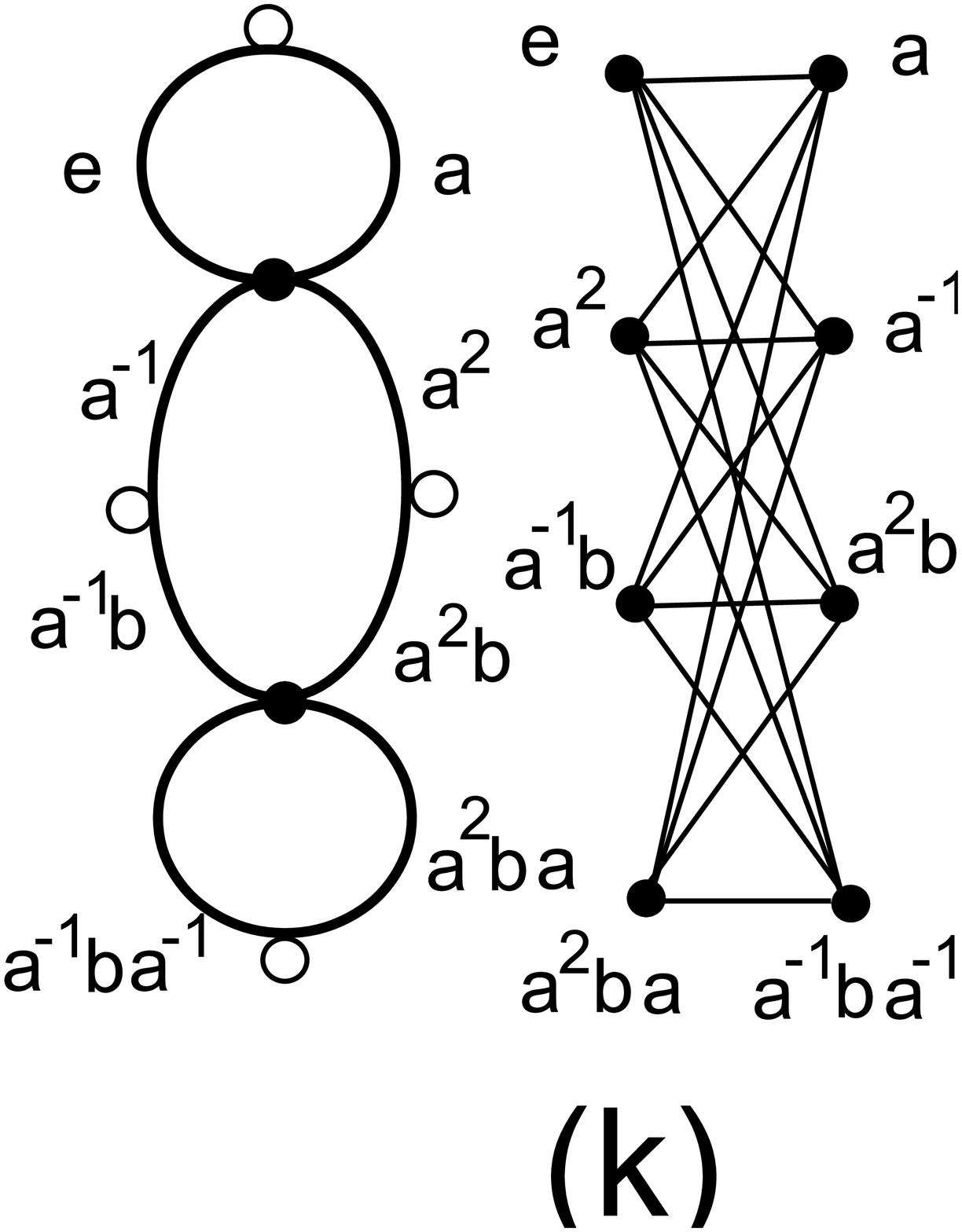}
\caption{(i) A non-contextual hypermap (left) stabilizing the bipartite graph $K(3,3)$ (right) with the corresponding coset labelling. (j) A non-contextual map/dessin (top) stabilizing the Fano plane (bottom). (k) A non-contextual map/dessin (left) stabilizing the bipartite graph $K(4,4)$ (right) with the corresponding coset labelling. The graph $K(3,3)$ and the Fano plane are given a set of two- and three-qubit coordinates, respectively. The seven $3$-qubit coordinates in the Fano plane are mutually commuting and the product of the three coordinates on any line is the identity matrix (see Sec. 4).}
 \label{fig2}
\end{figure} 

Besides $P_1$, there are four permutation groups $P$ isomorphic to the group $\mathbb{Z}_2 \times \mathbb{Z}_6$ that may be used to recover $\mathcal{O}$. 
Two of them are tree-like as is the dessin shown in \cite[Fig. 2]{Dessins2013} and one can check that they are of the contextual type. The other two are non-contextual as the one shown in Fig. 1 (j) where the permutation group is
 $P_2=\left\langle  (1,2,4,6,5,3), (2,3)(4,5)\right\rangle$ [here the $G$-set $\{1,2,3,4,5,6\}$ is an ordered set of indices for the transversal $\{e,a,a^{-1},a^2,a^{-2},a^3\}$]. The Belyi curve for this dessin is easily found as 
$$f_2(x)=\frac{4}{27}\frac{x^6}{(x^2-1)^2},$$
and this function allows to coordinatize the vertices of the dessin in Fig. 2 (j).

Observe that all coordinates of both dessins in Fig. 1 live in the extension field $\mathbb{Q}(\sqrt{3})$.

\subsection*{Recovering the n-simplex and the $2r$-ortoplex}
\noindent

In our earlier work, we found that there exist many dessins $\mathcal{D}$ of index $n$ stabilizing the $n$-simplex and, when $n$ is even ($n=2r$), also stabilizing the $r$-orthoplex: the smallest orthoplex structures are the square, the octahedron and the $16$-cell for which $r=2,3$ and $4$, respectively \cite[Table 1]{Dessins2014}. The vertices in these {\it trivial} structures can always be coordinatized in terms of single-generator cosets $a^q$, for some $q \in \mathbb{Z}$ and $|q|\le n/2$. The commutator of every pair of cosets is thus the identity so that the $n$-simplex and the $r$-orthoplex are of the non-contextual type. 

\subsection*{Recovering geometries of small index}
\noindent

In this line of thoughts, the first connected and non-trivial geometries that are $\mathcal{D}$-stabilizable are the bipartite graph $K(3,3)$, the Fano plane and the bipartite graph $K(4,4)$ corresponding to the index (the number of vertices) $6$, $7$ and $8$, respectively.

 As for the graph $K(3,3)$ and stricto sensu, we found no dessin $\mathcal{D}$ built from the free group $G$ and that respects the non-contextual definition that the edges of the graph are defined by commuting cosets. But there exists an hypermap built on the general free group $F=\left\langle a,b \right\rangle$ that satisfies the latter constraint as shown on Fig. 2 (i).

	As for the Fano plane, there exists a map and a single-generator and non-contextual coset coordinatization shown in Fig. 2 (j). Then, a dessin d'enfant $\mathcal{D}$ stabilizing the bipartite graph $K(4,4)$ in terms of two-generator cosets and  in a non-contextual way is shown in Fig. 2 (k).

\begin{figure}
\centering 
\includegraphics[width=7cm]{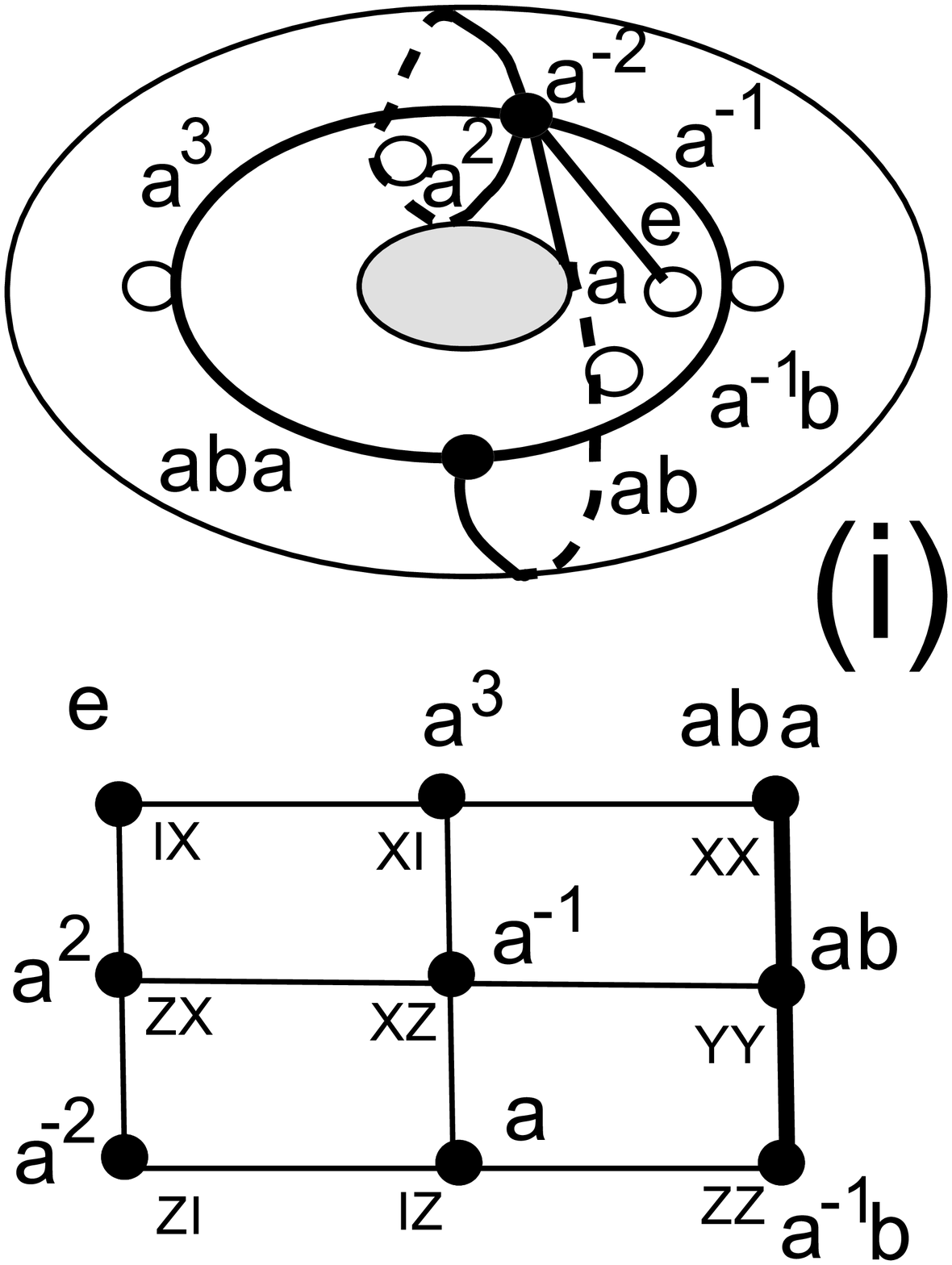}
\includegraphics[width=7cm]{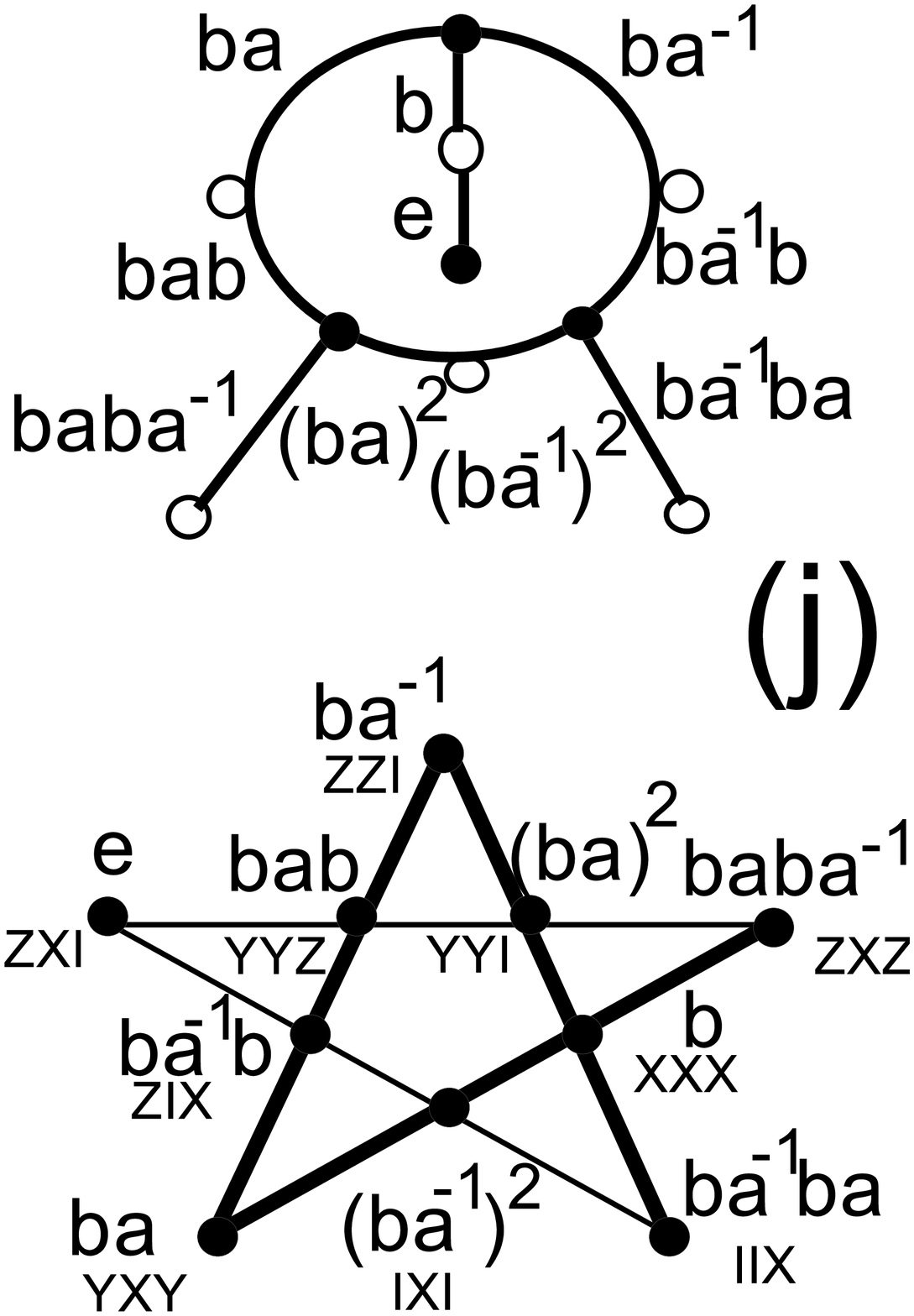}
\caption{(i) A contextual map/dessin (top) stabilizing the Mermin square (bottom) with the corresponding coset labelling. The right hand side column is defective as in the original proof of Kochen-Specker theorem derived for two-qubit coordinates. (j) A contextual map/dessin (top) stabilizing Mermin pentagram (bottom). The thick lines are defective: not all of their cosets are commuting. The lines of the pentagram are given three-qubit coordinates in such a way that the product of operators on a thick line is minus the identity matrix, see \cite{Planat13} for all such configurations.}
\label{fig4}
\end{figure}

\subsection*{Recovering contextual geometries}
\noindent

The smallest size and contextual point-line geometry (in the sense of our definition in Sec. 2) is the $(3 \times 3)$-grid, also known as Mermin's square in honor of D. Mermin who made use of it to prove the Kochen-Specker theorem in the four-dimensional Hilbert space \cite{Mermin1993,Planat2013bis}. There exists a unique (genus 1) map stabilizing the Mermin's square shown in Fig. 3 (i) (also pictured in \cite[Fig. 7]{Dessins2013}). There are two subgroups $S_1\cong \mathbb{Z}_1$ (a single-element group) and  $S_2\cong \mathbb{Z}_2$ (a two-element group)  stabilizing a pair of elements in the permutation group attached to the dessin. Both stabilizer subgroups lead to Mermin squares that are skewed to each other. The former one is non-contextual that is the cosets on the lines/triads of the grid are commuting (not shown); the other grid is contextual as shown at the bottom of Fig. 3 (i) in that the right column does not had all his triples of cosets commuting. This observation establishes a striking parallel with the proof of the Kochen-Specker theorem based on this geometry. It does not come as a surprise that two other $\mathcal{D}$-stabilized geometries of index $9$, the Pappus and Hesse configurations, that contain the $(3 \times 3)$-grid, are also found to be contextual. But the multipartite graph $K(3,3,3)$ admits a two-generator non-contextual dessin (not shown).

For the index ten, there are four non-trivial graph/configurations that may be $\mathcal{D}$-stabilized. The Petersen graph and the Mermin's pentagram corresponds to two distinct stabilizer subgroups $S_i$, ($i=1..2$) of the permutation group $P$ for the relevant $\mathcal{D}$, see \cite[Fig. 10]{Dessins2014}. Another disguise of both is the Desargues configuration that can also be $\mathcal{D}$-stabilized as shown in \cite[Fig. 11]{Dessins2014}. The last $\mathcal{D}$-stabilized connected structure is the bipartite graph $K(5,5)$. All the four structures are contextual. In Fig. 3 (j), one plots one of the three $\mathcal{D}$-stabilized pentagrams. In the corresponding dessin, one notices that coordinates on the right of the vertical axis are obtained from the ones at the left by replacing $a$ by $a^{-1}$. The bold lines of the pentagram are those that are defective for the commutativity of the cosets on them.

\section{Contextuality in maximum sets of commuting observables}
\noindent
 
While maximum sets of mutually commuting observables arising from the two-qubit and three-qubit Pauli group are non-contextual -they correspond to the triangle and Fano plane \cite{Levay2008} [also Fig. 2 (j)], respectively- this is no longer true for commuting sets in the general $n$-qubit Pauli group, $n>3$.

\begin{figure}[ht]
\centering 
\includegraphics[width=7cm]{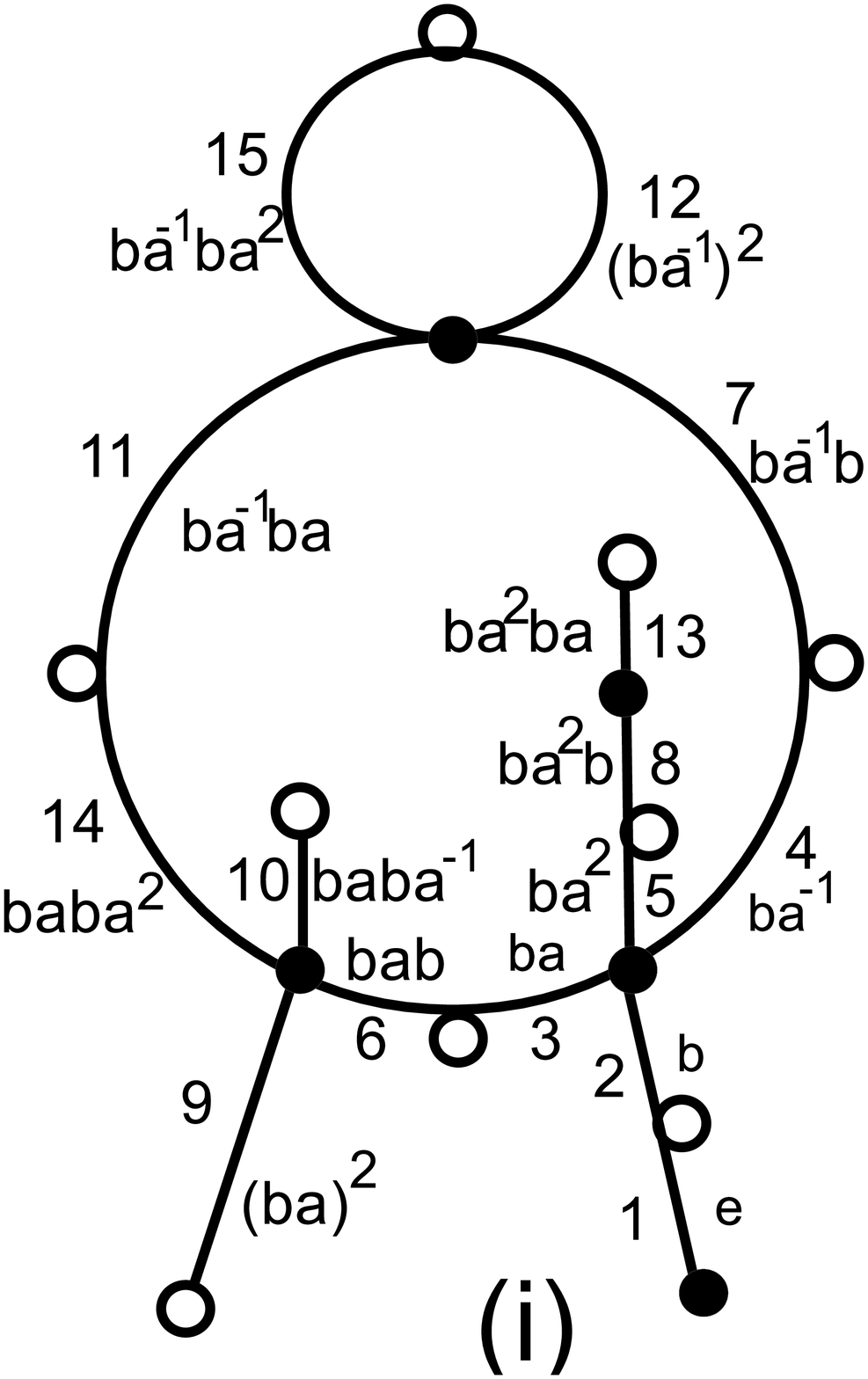}
\includegraphics[width=8cm]{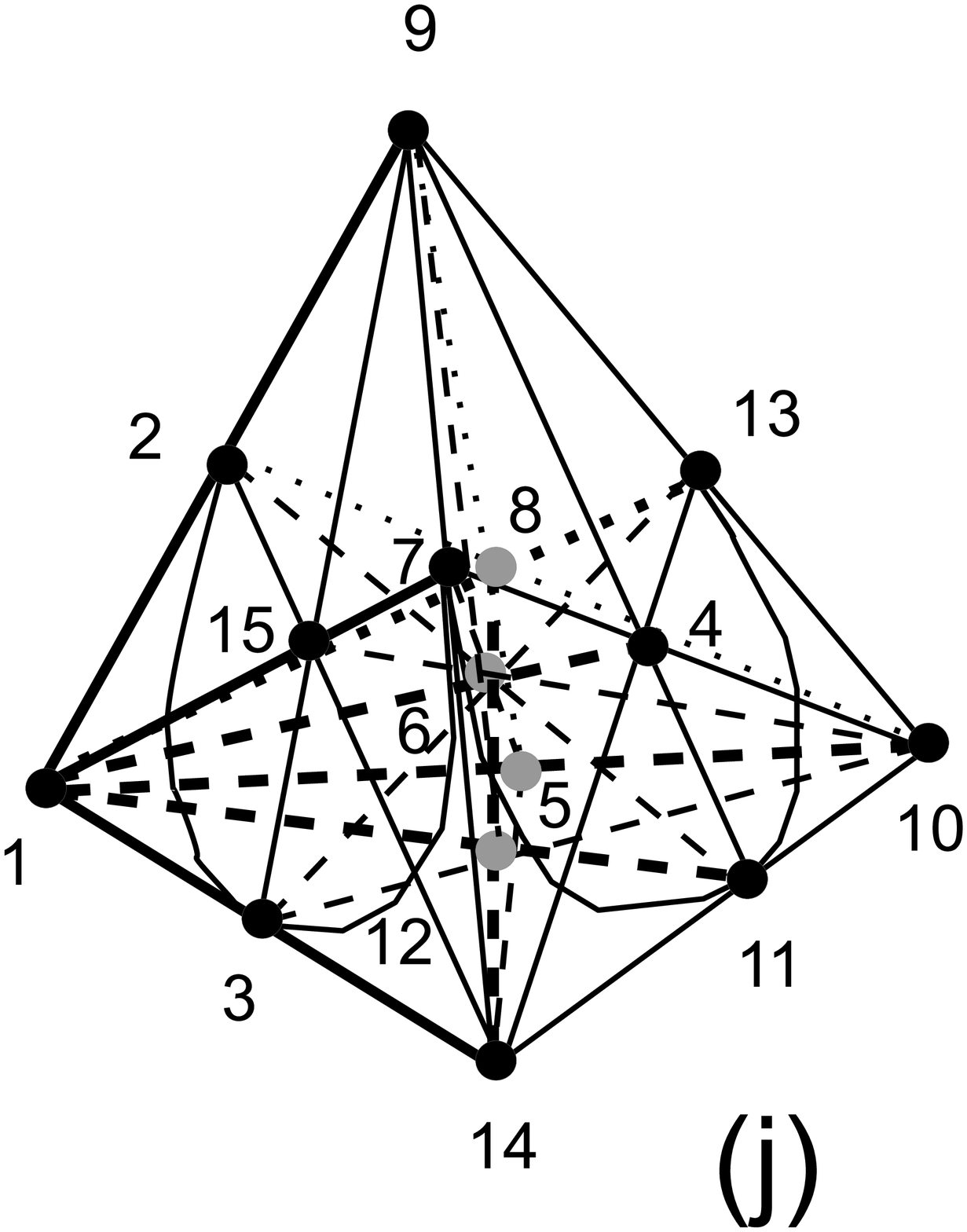}
\caption{ (i) A contextual hypermap/dessin stabilizing the projective space $PG(3,2)$. The edges are labelled in terms of an indexed $G$-set and the corresponding cosets. (ii)  The projective space $PG(3,2)$ as a model of a $4$-qubit maximal commuting set. A few lines/triads inside are not drawn. Thick triads are such that their points are commuting cosets. \label{fig3}}
\end{figure}

In the four-qubit Pauli group, such a maximum set comprises $2^4-1=15$ operators arranged as $35$ triads on which the product of operators is the identity matrix. The point/line geometry is that of the projective space $PG(3,2)$. We ask two questions:
(i) does it exist a dessin of index $15$ stabilizing $PG(3,2)$? (ii) are the coset coordinates such that each line of $PG(3,2)$ has commuting cosets as its points? The answer to (i) is yes but the answer to (ii) is definitely no, as shown below. 

For recovering/stabilizing $PG(3,2)$, we start from a subgroup $G'$ of the free group $F$ of finite representation $G'=F/[b^2=a^8=(ba^{-1})^7]=1$. The selected relations at the quotient were suggested by the finite representation of the symmetric group $S_8$. There are four subgroups $H$ of $G'$ of index $15$ and permutation group $P$ isomorphic to the alternating group $A_7$, of cardinality $2520$. The stabilizer of one point in $P$ is the group $PSL(2,7)$ and that of a pair of points is the alternating group $A_4$. The geometry that arises from $A_4$ is that of the projective space $PG(3,2)$. We selected the dessin that produces as many lines as possible such that their points/cosets satisfy the commutation law, only nine over the $35$ have this property. Thus the three-dimensional projective space is clearly contextual. The result is illustrated in Fig. 4 with the dessin (i) and the corresponding three-dimensional projective space (j).

A similar methodology holds for $PG(n,2)$, $n>3$, which is a model of a maximal commuting set in the $(n+1)$-qubit setting. For the general case, one finds that $PG(n,2)$ contains at least a copy of $PG(n-1,2)$ [a perp-set of $PG(n,2)$] as a non-contextual subset.

\begin{figure}[ht]
\centering 
\includegraphics[width=8cm]{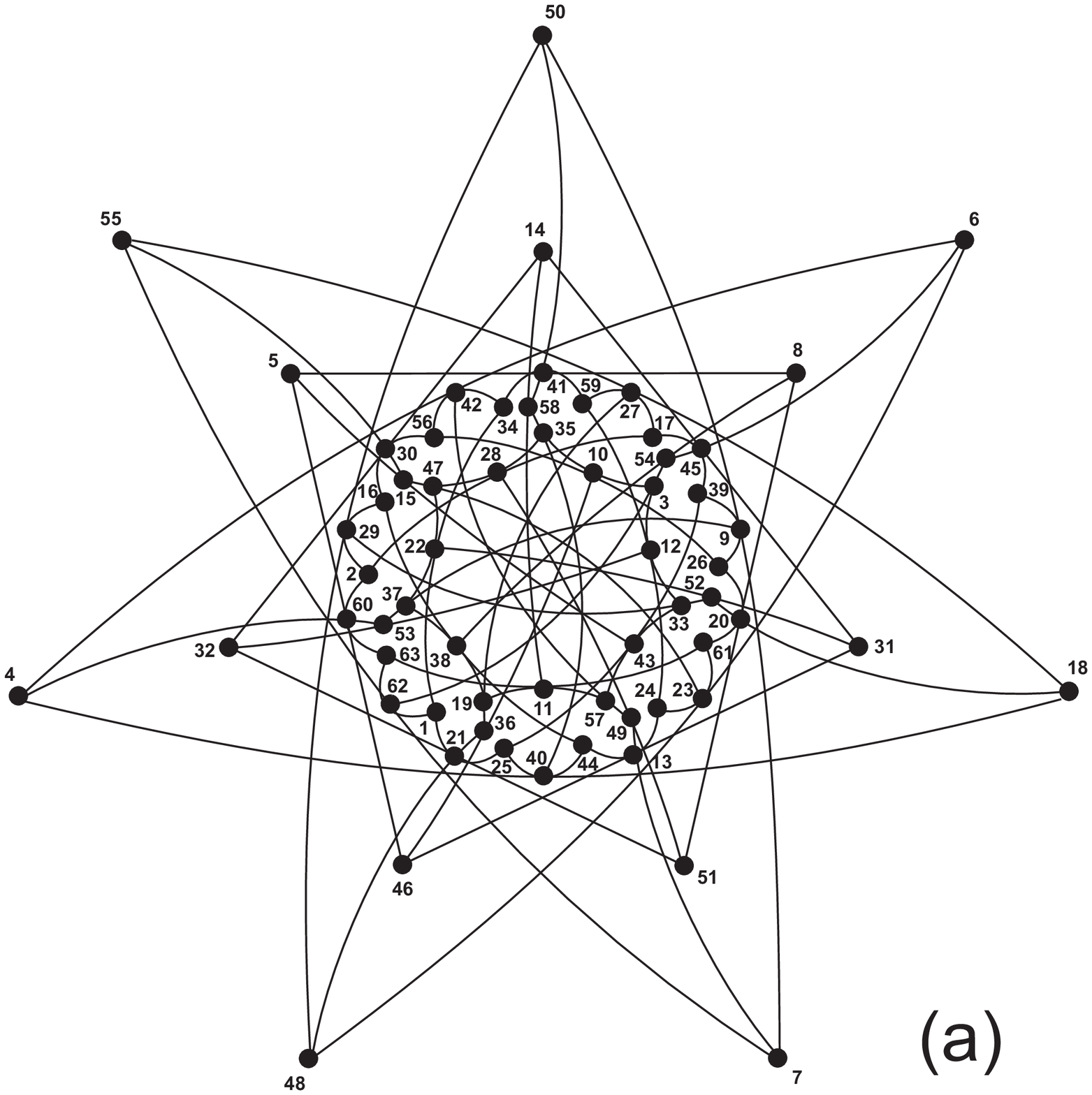}
\includegraphics[width=6cm]{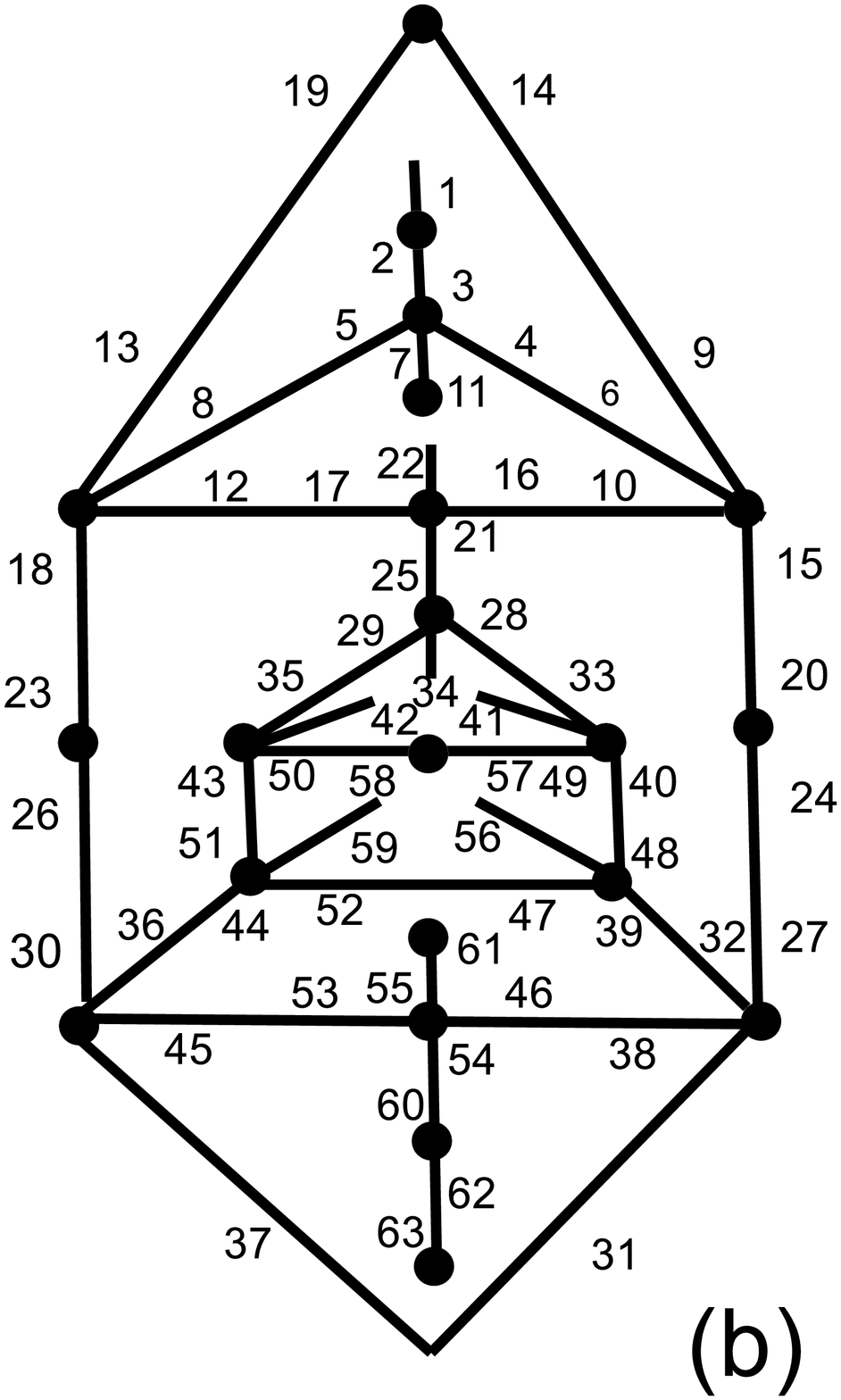}
\caption{The generalized hexagon $GH(2,2) (a)$ is stabilized by the genus zero dessin (b). Only three lines of $GH(2,2)$ have mutually commuting cosets (not shown).}
\end{figure}

\begin{figure}[ht]
\centering 
\includegraphics[width=8cm]{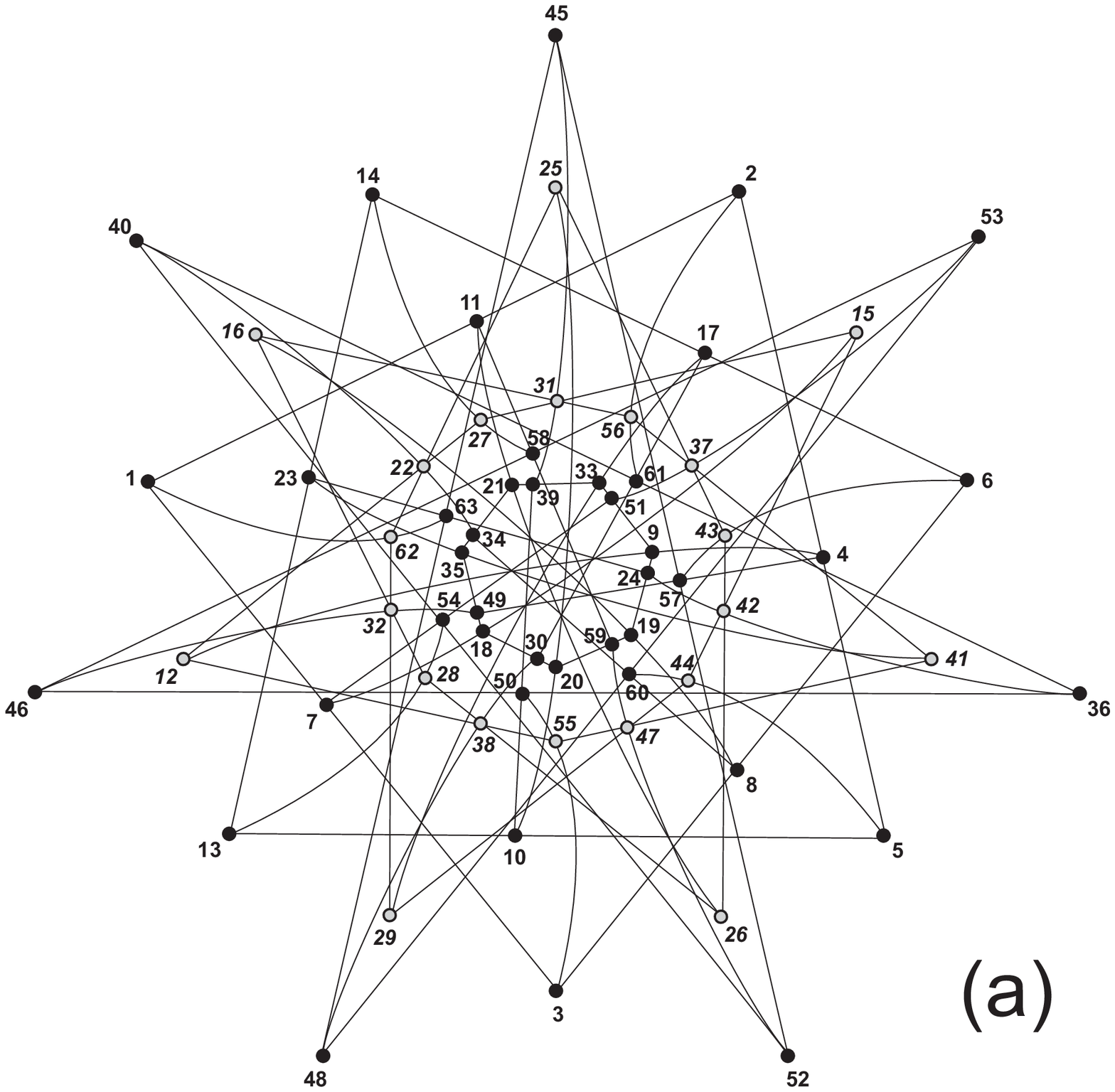}
\includegraphics[width=6cm]{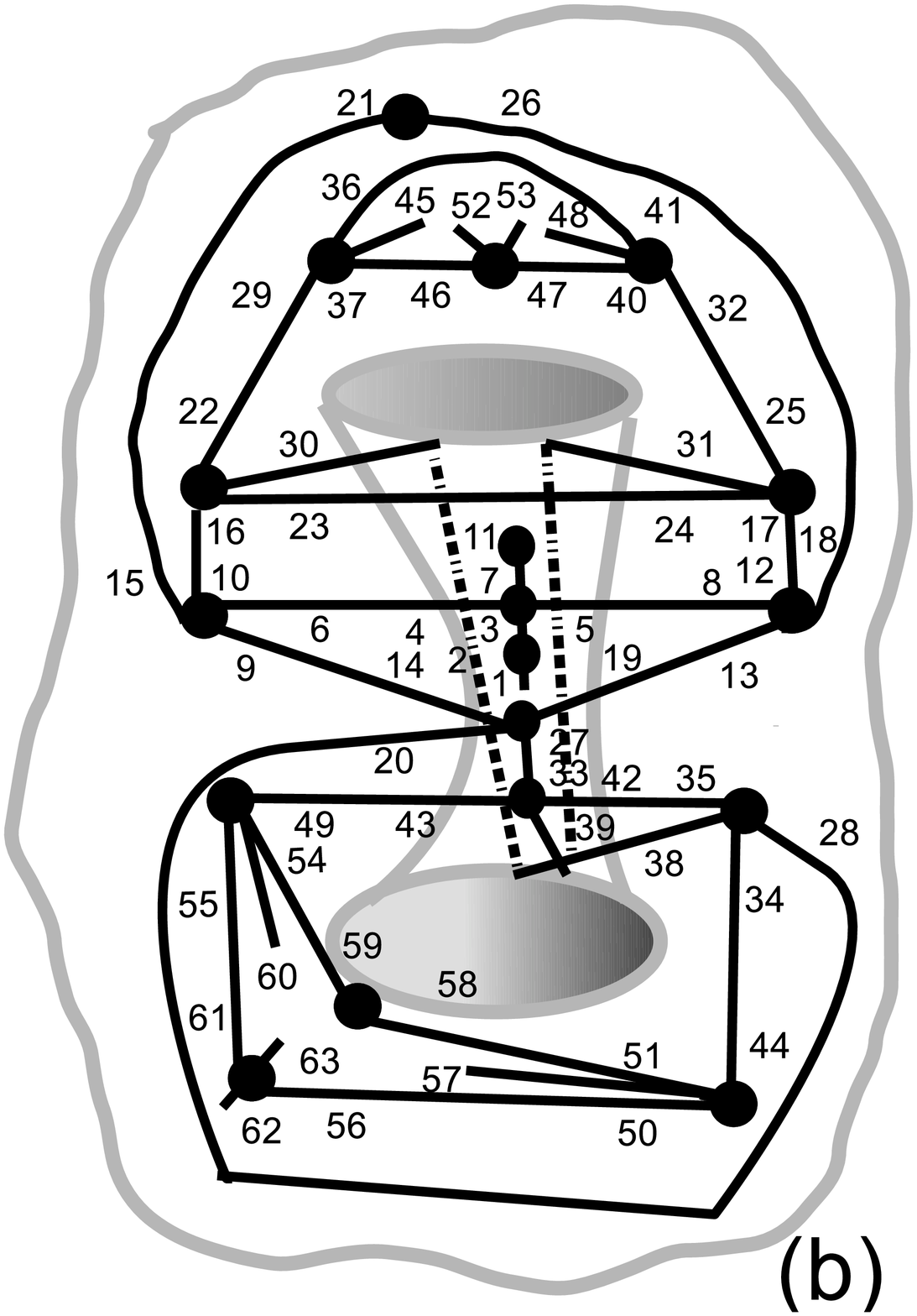}
\caption{The dual of generalized hexagon $GH(2,2) (a)$ is stabilized by the genus one dessin (b). Only four lines in this dual have mutually commuting cosets (not shown).}
\end{figure}

\section{Quantifying geometrical contextuality}
\noindent

Once one accepts that the coset structure of finite geometries induced by dessins d'enfants may reveal a geometrical contextuality, it is tempting to arrive at a quantification of such a contextuality. Quantifying quantum contextuality is currently an active subject \cite{Abramsky2011,Grudka2014,Durham2014}. From our definition in Sec. 2, a contextual finite geometry $\mathcal{G}$ cannot have all its lines encoded with commuting cosets. Let $l$ be the number of lines of $\mathcal{G}$ and $u$ the number of them with commuting cosets. Thus $\mathcal{G}$ is contextual as soon as $\frac{l}{u}>1$ so that a possible measure of contextuality is $c=\frac{l-u}{l}$ where $0 \le c \le 1$ and $c$ vanishes for a non-contextual    geometry.

Our earlier work featured a few generalized polygons [the Mermin's square $GQ(2,1)$ is the smallest one] \footnote{ A Tits generalized polygon (or generalized $n$-gon) is a point-line incidence structure whose incidence graph has diameter $n$ and girth $2n$. A generalized polygon of order $(s,t)$ is such that every line contains $s+1$ points and every point lies on $t+1$ lines. According to Feit-Higman theorem, the finite generalized $n$-gons, $s,t \ge 2$, exist for $n=2,3,4,6$ or $8$. One uses the notation $GQ$ (for a generalized quadrangle), $GH$ (for a generalized hexagon) and $GO$ (for a generalized octagon) corresponding to $n=4,6$ and $8$, respectively.} useful for encoding the commutation law of quantum operators in the generalized Pauli group. In such structures, the number $h$ of geometric hyperplanes is known \footnote {A geometric hyperplane of a generalized polygon is a proper subspace meeting each line at a unique point or containing the whole line. The set of hyperplanes can be constructed in an efficient way by using an addition law for the hyperplanes: the \lq sum' of two hyperplanes is just the complement of the symmetric difference in the relevant $G$-set of indices labelling the vertices of the geometry \cite{Frohard1994,GreenSaniga2013}.} and $h$ happens to grow with the contextual parameter $\frac{l}{u}$ roughly as $\log_2 h$, as shown in Table 1.

\begin{table}[ht]
\begin{center}
\caption{Geometric contextuality measure $l/u$ ($l$ the number of lines and $u$ the number of them with commuting cosets) for a few generalized polygons compared to the base-two logarithm of the number $h$ of hyperplanes within the selected geometry.}
\begin{tabular}{||l|crcl|r|}
\hline \hline
Geometry  & l & u  & l/u &   $\log_2(h)$ & Remark\\
\hline\hline
$GQ(2,1$) &  6 &  5 & 1.2 &  4&  Mermin square, Fig. 3 (i)\\
$GQ(2,2)$ &  15 &  3 & 5 &  5&  two-qubit commutation \cite{Dessins2013,Dessins2014}\\
$GQ(2,4)$ &  45 &  5 & 9 &  6&  black-hole/qubit analogy \cite{Dessins2013}\\
$GH(2,1)$ &  14 &  2 & 7 &  8&  in the dual of $GH(2,2)$ \cite{Frohard1994}\\
$GO(2,1$) &  30 &  2 & 15 &  16&  in $GO(2,4)$ \cite{DeBruyn2011}\\
$GH(2,2$) &  63 &  3 & {\bf 21} &  14&  \cite{Frohard1994} and Fig. 5\\
dual of $GH(2,2$) &  63 &  4 & {\bf 15.75} &  14&  \cite{Frohard1994} and Fig. 6\\
\hline
\hline
\end{tabular}
\end{center}
\end{table}

\subsection*{The generalized hexagon $GH(2,2)$}
\noindent

The generalized hexagon $GH(2,2)$ (with $63$ vertices and dually $63$ lines/triads) is an excellent geometrical model of $3$-qubit contextuality \cite{Levay2008,Planat13}. The hexagon $GH(2,2)$ and its dual can be stabilized by dessins d'enfant. For recovering them, one can start from a subgroup $G''=F/[b^2=a^4=(ab)^7=(a,b)^6]$ of the free group $F$. There are just two subgroups $H$ of $G''$ of index $63$ inducing a dessin with permutation group $P$ of order $12096$. The first dessin in Fig. 5 (b) is of genus $0$  and signature $(B,W,F,g)=(21,35,9,0)$, it stabilizes $GH(2,2)$ shown in Fig. 5 (a) through the stabilizer subgroup $S_1 \cong \mathcal{Z}_2^3 \rtimes \mathcal{Z}_2^2$. The second dessin in Fig. 6 (b) is of genus $1$ and signature $(B,W,F,g)=(18,36,9,1)$, it stabilizes the dual of  $GH(2,2)$ shown in Fig. 6 (a) through the stabilizer subgroup $S_2 \cong E_{32}^+$ (the extraspecial group of order $32$).

 It has been recognized that the size $12096$ of the automorphism group of $GH(2,2)$ is also the number of $3$-qubit pentagrams and is related to the number of copies of hyperplanes in each class \cite{Planat13}. For this hexagon (resp. its dual), the geometrical contextuality measure $\frac{l}{u}=21$ (resp $\frac{l}{u}=15.75$) is larger than $\log_2 h=14$, while it is not the case for the other polygons in the table. The hexagon $GH(2,2)$ can be consided as \lq strongly contextual' in this respect.

\section{Conclusion}
We provided a striking comparison between the commutativity of multiple qubit quantum observables and that of cosets of subgroups of the two-generator free group. This parallel allowed us to propose a new definition of contextuality based on the coset structure of Grothendieck's dessins d'enfants, in close correspondence with the standard quantum one. In particular, geometric contextuality in small generalized polygons starting with the $(3 \times 3$)-grid was investigated. Further work may focus on identifying and quantifying contextuality in higher size geometries. Since a complex algebraic curve defined over the field $\bar{\mathbb{Q}}$ of algebraic numbers [that is a Belyi function $f(x)$] \cite{Groth84,Lando2004} is attached to any dessin d'enfant $\mathcal{D}$, it is expected that the contextuality criterion features specific curves through the action of the group of automorphisms $\mbox{Gal}(\bar{\mathbb{Q}}/\mathbb{Q})$ of the field $\bar{\mathbb{Q}}$ (the absolute Galois group) on dessins that would be helpful to recognize. We can say more about Nature than in Bohr's century. The coset language, with only two letters $a$ and $b$, is enough for quantum contextuality. Even the sporadic groups enter this frame \cite{PlanatFQXi2015}.

\end{document}